%% file: main.tex
\documentclass[sigconf]{acmart}

\copyrightyear{2026}
\acmYear{2026}
\setcopyright{cc}
\setcctype{by}
\acmConference[WWW '26]{Proceedings of the ACM Web Conference 2026}{April 13--17, 2026}{Dubai, United Arab Emirates}
\acmBooktitle{Proceedings of the ACM Web Conference 2026 (WWW '26), April 13--17, 2026, Dubai, United Arab Emirates}
\acmPrice{}
\acmDOI{10.1145/3774904.3792558}
\acmISBN{979-8-4007-2307-0/2026/04}

\usepackage{booktabs} 
\usepackage{array} 
\usepackage{multirow}
\usepackage{multicol}
\usepackage{fontawesome5}
\usepackage{cleveref}
\usepackage{enumitem}
\usepackage{subcaption}
\usepackage{geometry}
\usepackage{lipsum}

\usepackage{xcolor}

\usepackage{tikz}
\usepackage{amsmath}
\usetikzlibrary{decorations.pathreplacing, positioning}

\usepackage{pifont}
\usepackage[most]{tcolorbox}

\usepackage{url}

\definecolor{fg}{RGB}{34,139,34}
\newcommand*\colourcheck[1]{%
  \expandafter\newcommand\csname #1check\endcsname{\textcolor{#1}{\ding{52}}}%
}
\newcommand*\colourtimes[1]{%
  \expandafter\newcommand\csname #1times\endcsname{\textcolor{#1}{\ding{56}}}%
}
\colourcheck{fg}
\colourtimes{red}

\newif\ifcomment
\commentfalse
\ifcomment
\newcommand{\kg}[1]{\textcolor{teal}{\textbf{Krishna: #1}}}
\newcommand{\sz}[1]{{\textcolor{violet}{\textbf{Savvas: #1}}}}
\newcommand{\sep}[1]{\textcolor{red}{\textbf{Sep: #1}}}
\newcommand{\ad}[1]{\textcolor{cyan}{\textbf{Abhisek: #1}}}

\else
\newcommand{\sz}[1]{}
\newcommand{\sep}[1]{}
\newcommand{\ad}[1]{}
\newcommand{\kg}[1]{}

\fi

\renewcommand\footnotetextcopyrightpermission[1]{} 
\makeatletter
\def\@copyrightspace{\relax}
\makeatother

\makeatletter
\def\@copyrightspace{\relax}
\makeatother

\begin{document}

\title[Does Ad-Free Mean Less Data Collection?]{Does Ad-Free Mean Less Data Collection? An Empirical Study of Platform Data Practices and User Expectations}

\author{Sepehr Mousavi}
\orcid{0000-0002-3023-4479}
\affiliation{%
  \institution{Max Planck Institute for Software Systems}
  \city{Saarbrücken}
  \country{Germany}}
\email{smousavi@mpi-sws.org}

\author{Abhisek Dash}
\orcid{0000-0002-5300-8757}
\affiliation{%
  \institution{Max Planck Institute for Software Systems}
  \city{Saarbrücken}
  \country{Germany}}
\email{adash@mpi-sws.org}

\author{Savvas Zannettou}
\orcid{0000-0001-5711-1404}
\affiliation{%
  \institution{TU Delft}
  \city{Delft}
  \country{Netherlands}}
\email{s.zannettou@tudelft.nl}

\author{Krishna P. Gummadi}
\orcid{0000-0003-1256-8800}
\affiliation{%
  \institution{Max Planck Institute for Software Systems}
  \city{Saarbrücken}
  \country{Germany}}
\email{gummadi@mpi-sws.org}

\renewcommand{\shortauthors}{Sepehr Mousavi, Abhisek Dash, Savvas Zannettou, and Krishna P. Gummadi}

\input{abstract}


\maketitle

\newcommand\webconfavailabilityurl{https://github.com/sepehrmousavi/subscription-data-practices}
\ifdefempty{\webconfavailabilityurl}{}{
\begingroup\small\noindent\raggedright\textbf{Resource Availability:}\\
The source code of this paper has been made publicly available at \url{\webconfavailabilityurl}.
\endgroup
}

\newpage
\input{introduction}
\input{related_work}
\input{rq1}
\input{rq2}
\input{discussion}

\bibliographystyle{ACM-Reference-Format}
\bibliography{references}

\input{appendix}

\end{document}

%% file: abstract.tex
\begin{abstract}
Online platforms increasingly offer "paid" \textit{ad-free subscriptions} as an alternative to the traditional "free" \textit{ad-based model}.
The transition to ad-free models ostensibly removes advertising as a key justification for data processing under the GDPR.
So, normatively, platforms should collect less user data.
However, platforms may justify continued data collection as a means to provide an improved, personalized experience.
This tension between privacy principles and platform incentives raises a critical underexplored question:
do data collection practices vary between ad-free and ad-based subscription models?

In this paper, we shed light on this important privacy issue by investigating the alignment between platform data collection practices and related user expectations.
With respect to data collection process, our analyses of data exports from three major online platforms — Instagram, Facebook, and X — reveal that these platforms continue to retain or collect some ad-related data, even in ad-free subscriptions.
With respect to user expectations, our survey among 255 participants on Prolific reveals that 69\% of the participants normatively expect data collection to be reduced, indicating their expectation of improved digital privacy in an ad-free model.
However, when asked what they think actually happens, 63\% of these participants believed that platforms would still collect about the same amount of data, highlighting skepticism about platform practices.
Our findings not only indicate a significant disconnect between data practices and normative user expectations, but also raise serious questions about platform compliance with core GDPR principles, such as purpose limitation, data minimization, and transparency.
\footnote{\textcolor{red}{This paper has been accepted at The ACM Web Conference 2026. Please cite the version appearing in the conference proceedings.}}
\end{abstract}

%% file: introduction.tex
\section{Introduction}~\label{Sec: Intro}

\noindent
Online platforms such as Instagram, Facebook, and X (formerly Twitter) are increasingly offering users a choice to access their services in two structurally different models, namely ~\textit{ad-based} and ~\textit{ad-free}.
In the ad-based model, platforms generate revenue from advertisers by serving ads to their users, who receive free access to their services.
In contrast, the ad-free model allows platforms to generate revenue directly from their users in the form of subscription fees, without interrupting their experience with ads. 
Put differently, in ad-free service models, users are the \textit{customers} of the platform services, while in ad-based service models, users are the \textit{commodities} offered by the platforms to their advertising customers.

In this paper, we focus on \textit{comparing the data gathering practices} of very large online platforms under these two service models.
While the privacy research community has extensively examined various aspects of data collection by online platforms, little attention has been given to how data collection practices change when switching between the two models.

Comparing a platform's data collection practices under ad-based and ad-free models is important both from a regulatory as well as an end-user perspective.
From a regulatory perspective, the General Data Protection Regulation (GDPR) mandates a set of core principles for data processing, such as purpose limitation and data minimization (Article 5 (1))~\cite{gdpr}.
In both models, platforms have an incentive to collect user data to provide personalized services to users.
However, since serving targeted ads is no longer a purpose in the ad-free model, platforms should collect less user data than in the ad-based model to adhere to the purpose limitation and data minimization principles.
Furthermore, in practice, the lack of a clear distinction between personal data for targeted advertising and service personalization may allow platforms to justify any data collected based on their legitimate interest (Article 6 (1)(f))~\cite{gdpr}.

Beyond investigating platform practices, understanding user expectations is equally important as they impact how users choose between the two service models. 
Ascertaining users' normative expectations (i.e., what they believe platforms \textit{should} do), with respect to data gathered about them, would reveal whether users expect better privacy in an ad-free model, i.e., \textit{whether their willingness to pay subscription fees is to avoid ads or is based on expectations of protecting their privacy.}
Similarly, assessing their descriptive expectations (i.e., what they believe platforms \textit{will} do) may reveal doubts about whether platforms will meet users’ normative expectations, which could affect the adoption of ad-free models.
Finally, checking user expectations against actual platform practices would help users make more informed service choices, as they can validate their expectations or correct their misperceptions.

\smallskip
\noindent
\textbf{Current work: }
Against this background, in this work, we attempt to answer the following crucial questions:

\smallskip
\noindent
\textbf{RQ1}: \textit{How do platforms change their data collection practices in an ad-free model as compared to an ad-based model?}
\\
\textbf{RQ2~(a)}: \textit{What are end-users' normative and descriptive expectations of personal data collection practices adopted by platforms?}
\\
\textbf{RQ2~(b)}: \textit{How do these practices affect users' choice of service model?}

To answer the first question, we collected data from three major online platforms that provide their services in ad-based and ad-free subscription models: Instagram, Facebook, and X.
First, we created accounts on these platforms in their ad-based subscription model and consumed content online. 
Then, we switched to an ad-free model for one month and continued to use the services. 
Next, we exercised our \textit{Right of access} (GDPR Article 15(3)) to personal data by requesting a copy of all the data gathered by the platforms. 
We compared the collected data exports during ad-based subscription and ad-free subscription to investigate the differences in data collection practices. 

To answer the second question (collectively), we surveyed $255$ participants on the online crowd-sourcing platform Prolific.\footnote{\url{https://prolific.com}}
We asked these participants about their normative expectation (\textit{should} platforms collect less data in ad-free subscription) and descriptive expectation (\textit{would} platforms collect less data in ad-free subscription) on data collection by platforms across different subscription models.
Additionally, we also asked our participants if platforms continuing a similar amount of data collection in the ad-free model (as compared to the ad-based model) would affect their decision to choose an ad-free model. 

\smallskip
\noindent
\textbf{Main findings:}
Next, we summarize the answers to our RQs.

\noindent
\underline{Answer to RQ1:}
Despite removing some ad-related data under the ad-free model, all platforms continue to retain or collect certain personal data used exclusively for advertising purposes.

\noindent
$\bullet$
~Instagram and Facebook remove some ad-related data once users subscribe to the ad-free model.

\noindent
$\bullet$
~However, under the ad-free model, all three platforms continue to retain or collect some personal data that are exclusively used for advertising purposes.

\noindent
\underline{Answer to RQ2:}
We observe a significant gap between the normative and descriptive expectations of our survey participants. 

\noindent
$\bullet$ Majority (69\%) of participants normatively expect less data should be collected in the ad-free model, indicating they expect better digital privacy in the ad-free model.

\noindent
$\bullet$ However, only 17\% of participants descriptively expect less data to be collected in the ad-free model.

\noindent
$\bullet$ Finally, 34\% of participants mentioned that the continued collection of a similar amount of data would deter them from switching to an ad-free subscription model.

\noindent 
\textbf{Implications:}
In summary, our findings reveal a misalignment between user expectations, platform practices, and regulatory requirements.
Users are \textit{not} getting the privacy they expect, even in ad-free models.
Platforms often fall short of GDPR obligations, and potential poor coordination between legal, engineering, and product teams may prevent the implementation of compliant data practices.
These gaps point to the potential role of clearer regulatory guidance and better coordination across stakeholders.

%% file: related_work.tex
\section{Background \& Related Work}

In this section, we contextualize our work by (a)~setting the necessary legal background, (b)~highlighting some of the key non-compliance of platforms with regulations, and (c)~reporting prior research in computer science related to the current work.

\subsection{Legal Background}

While several provisions of the General Data Protection Regulation (GDPR)~\cite{gdpr}, Digital Services Act (DSA)~\cite{dsa}, and Digital Markets Act (DMA)~\cite{dma} are broadly relevant, covering all of them is beyond the scope of this work.
Therefore, we focus on the GDPR provisions that are most pertinent to our study.

\smallskip
\noindent
\textbf{Principles of data processing:}
The processing of personal data by online platforms within the European Union (EU) is primarily governed by the GDPR, which establishes foundational \textit{principles} of data processing.
Central to these principles is Article 5, mandating that personal data be processed lawfully, fairly, and transparently (Article 5(1)(a)). 
Furthermore, purpose limitation (Article 5(1)(b)) requires data to be collected for specified, explicit, and legitimate purposes and not further processed in a manner incompatible with those purposes. 
Finally, data minimization (Article 5(1)(c)) dictates that data collection must be adequate, relevant, and limited to what is necessary in relation to the purposes for which it is processed.

\smallskip
\noindent
\textbf{Lawfulness of data processing:}
Article 6 of the GDPR mandates the core principles of data processing by listing valid legal bases.
Among these bases, \textit{legitimate interests} (Article 6(1)(f)) is the most pertinent to platform operations concerning advertising and personalization.
Platforms might invoke \textit{legitimate interests} to justify ongoing data collection for service personalization even in ad-free setting. 
However, such justification requires a careful three-part assessment: identifying a legitimate interest, ensuring the processing is necessary to achieve it, and critically balancing this against the fundamental rights and freedoms of the data subject -- a balance that, we posit, \textit{may shift when users are directly paying for a service}.

\smallskip
\noindent
\textbf{Rights of access by users:}
The GDPR also enables users to exercise rights to verify whether the principles and lawfulness of data processing are upheld by platforms.
To this end, a cornerstone for such user empowerment is the \textit{Right of Access} (Article 15). 
Specifically, Article 15(3) grants the right to obtain a copy of the personal data undergoing processing, enabling users and researchers alike to scrutinize actual data being processed.

\subsection{Violations by Tech Platforms}
\label{subsec:related_work_violations}

Major technology platforms have repeatedly faced regulatory actions in the EU, highlighting systemic concerns over data practices, particularly in advertising and personalization.
For instance, as a violation of the GDPR, Google was fined in 2019 for not properly disclosing to users about its data collection and processing practices in the context of personalized ads~\cite{nytimes_google}.
Amazon was also fined by the data protection body of Luxembourg for manipulating user consent in their advertising practices~\cite{reuters_amazon}.
These cases indicate a robust regulatory environment in which personal data collection for profiling, personalization, and advertising remains under constant scrutiny, regardless of the service model.

Meta’s introduction of a `consent or pay' model for Facebook and Instagram in the EU highlights the tension between data-driven business models and privacy regulation~\cite{meta_ad_free}.
Increasingly adopted by platforms~\cite{morel2023legitimate, rasaii2023thou}, this model offers users a choice between consenting to extensive data tracking used for personalization and advertising~\cite{bujlow2017survey, ada2022context}, or paying to limit such processing.
However, Data Protection Authorities questioned whether consent obtained under this scheme is freely given as required by GDPR Article 7, noting that subscription fees may unduly pressure users.
In response, the European Data Protection Board emphasized that valid `consent or pay' models must ensure a genuinely voluntary and informed choice~\cite{edpb_consent_or_pay}.
In 2025, the European Commission fined Meta for non-compliance of this model with GDPR and DMA consent requirements~\cite{eu_commission_fines_meta, gdpr, dma}.

\subsection{Related Work}

We contextualize our work within two strands of prior research: (1) user perception of data collection and targeted advertising, and (2) platform audits using GDPR data exports.

\smallskip
\noindent
\textbf{User perception of data collection and targeted advertising:}
Users recognize content personalization relies on data collection and view online tracking as a way for platforms to improve their services~\cite{kozyreva2021public, chanchary2015user}.
Moreover, they often view targeted advertising as beneficial for discovering relevant content and enhancing their overall experience~\cite{matic2017omg, sharma2023user}.
At the same time, concerns persist about how personal data are collected and used.
For instance, many users perceive targeted advertising as privacy-invasive or reliant on deceptive design practices~\cite{eslami2018communicating, mcdonald2010americans, ur2012smart, weinshel2019oh, carrascosa2015always, kyi2023investigating}.
Furthermore, users believe targeted advertising can contribute to the spread of harmful content, among other issues~\cite{li2012knowing, sharma2023user, subramani2020push, zeng2021makes}.
Our work builds on this line of research by studying users' normative and descriptive expectations of platform data collection across subscription models.

\smallskip
\noindent
\textbf{Platform audits using GDPR data exports:}
Researchers have been leveraging GDPR Right of Access to recruit participants who voluntarily donate their data exports.
Using these donated GDPR data exports, prior works have conducted research on data export reliability~\cite{karnam2026GDPR}, ad targeting mechanisms~\cite{wei2020twitter}, recommendation algorithms~\cite{vombatkere2024tiktok}, user behavior analysis~\cite{zannettou2024analyzing}, and societal topics ranging from news and politics to personal health and safety~\cite{blassnig2023googling, 2023haimintegratin, hase2024can, tiktok_addiction, alsoubai2024profiling}. 
We contribute to this line of research by employing data exports to examine platform data collection practices across subscription models.

%% file: rq1.tex
\section{Platform Data Gathering Practices Across Subscription Models}
\label{sec:rq1}

In this section, we investigate data collection practices of online platforms across subscription models and present our findings.

\subsection{Data Collection Methodology}

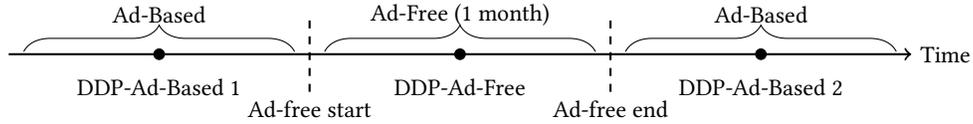
\begin{figure*}[t]
\centering
\begin{tikzpicture}

\draw[thick, ->] (0,0) -- (12,0) node[anchor=west] {Time};

\draw[dashed, thick] (4, -0.5) -- (4, 0.5);
\draw[dashed, thick] (8, -0.5) -- (8, 0.5);

\node[below] at (4, -0.5) {Ad-free start};
\node[below] at (8, -0.5) {Ad-free end};

\draw[decorate, decoration={brace, amplitude=10pt}, yshift=1pt]
  (0.2,0) -- (3.8,0) node[midway, yshift=14pt] {Ad-Based};

\draw[decorate, decoration={brace, amplitude=10pt}, yshift=1pt]
  (4.2,0) -- (7.8,0) node[midway, yshift=14pt] {Ad-Free (1 month)};

\draw[decorate, decoration={brace, amplitude=10pt}, yshift=1pt]
  (8.2,0) -- (11.8,0) node[midway, yshift=14pt] {Ad-Based};

\filldraw (2,0) circle (2pt) node[below=6pt] {DDP-Ad-Based 1};
\filldraw (6,0) circle (2pt) node[below=6pt] {DDP-Ad-Free};
\filldraw (10,0) circle (2pt) node[below=6pt] {DDP-Ad-Based 2};

\end{tikzpicture}
\caption{
Timeline of requested Data Download Packages (DDPs) across ad-based and ad-free subscription models of an online platform.
Initially, we start by obtaining an ad-based DDP (DDP-Ad-Based-1).
Then, we subscribe to the platform's ad-free model and request ad-free DDP (DDP-Ad-Free) throughout our subscription period.
Finally, we obtain a second ad-based DDP (DDP-Ad-Based 2) after our ad-free subscription ends and the account switches back to the ad-based model.
}
\label{fig:methodology}
\end{figure*}

To study data collection practices of online platforms across various subscription models, we need a method that allows us to obtain personal data that online platforms collect, store, and process.
This task is challenging because online platforms are generally non-transparent about their data collection practices, and the information disclosed in privacy policies is often incomplete. 

To overcome these challenges, we leverage the power of GDPR's Right of Access by the data subject (Article 15), which entitles individuals to obtain a copy of their data being held and processed by online platforms.
Platforms typically provide this information in the form of a Data Download Package (DDP).
These DDPs provide a comprehensive view of what a platform collects, stores, and processes about the user who exercised their GDPR right.
Crucially, DDPs capture data across devices (e.g., web and mobile), as well as information gathered through third parties, offering an invaluable perspective that is otherwise inaccessible.
By grounding our methodology in the information that platforms are legally required to disclose, we ensure that our analysis relies on data they demonstrably hold, thereby avoiding ambiguities around whether particular data were collected or retained.
For these reasons, we argue that DDPs are a suitable and reliable source for systematically studying data collection practices.

Building on this foundation, we designed a procedure to obtain comparable data from both ad-based and ad-free subscription modes.
~\Cref{fig:methodology} depicts a schematic overview of the steps we took for collecting such data.
First, two co-authors of this work created accounts in the platforms' default ad-based model from within the European Union.
We then subscribed to the ad-free model for one month and continued to use the service under this model.
After the subscription expired, we returned to the default ad-based model.
At each step of this procedure, we obtained a DDP: an initial ad-based DDP (DDP-Ad-Based-1) before switching to the ad-free subscription, an ad-free DDP (DDP-Ad-Free) three weeks into the subscription period, and a second ad-based DDP (DDP-Ad-Based-2) one week after returning to the ad-based model.

\noindent
\textbf{Platforms in this work:}
We focus on Instagram, Facebook, and X (formerly Twitter) in our study.
We chose these online platforms for multiple reasons.
First and foremost, all these platforms provide ad-based and ad-free subscription models, which is  essential for our comparative analysis.
Second, they operate their own advertising infrastructures rather than relying on external providers, meaning that the data we examine is generated and processed within the platforms' own ecosystems.
This makes them particularly relevant for studying how subscription models affect data collection.
Furthermore, DSA designates these platforms as Very Large Online Platforms (VLOPs)~\cite{dsa}.
Therefore, they are subject to heightened regulatory obligations and scrutiny.

\subsection{Categorizing Files Contained in DDPs}
\label{rq1_categorizing_files_ddp}

The DDPs from Instagram, Facebook, and X contain a large number of files, covering a wide variety of personal data categories.
We provide a detailed characterization of the contents of these DDPs in~\Cref{subsec:appendix_ddp}, and refer interested readers to~\cite{karnam2026GDPR} for further discussion.
While comprehensive, the diversity of these files makes systematic analysis challenging, as not all data collection serves the same purpose.
For instance, some files reflect information used exclusively for advertising, others pertain to personalization or platform functionality, and others contain data relevant to both.
Distinguishing between these categories is essential for our analysis: without it, we cannot determine whether ad-free subscriptions meaningfully change the scope of advertising-related data collection or whether platforms continue to collect comparable information for other purposes.

To enable such comparisons, we systematically categorized each file in the DDPs according to its relationship to advertising.
Two co-authors independently categorized each file as follows:
(1) \textbf{Ad-only} (the file is exclusively related to advertising and is not used for other non-advertising purposes),
(2) \textbf{Non-ad} (the file is unrelated to advertising),
and 
(3) \textbf{Mixed} (the file contains data that can be used for both advertising and other non-advertising purposes, such as content personalization).
In cases of disagreement, a third co-author independently made the final decision.
The categorization relied on multiple sources of information, including (a) the platforms' privacy policies; (b) README or explanatory documents provided within the DDPs (only applicable to X); (c) the folder structure used to organize files; and (d) file descriptions or embedded metadata.
The three co-authors conducting this classification are experienced researchers familiar with DDPs.
In total, we categorized 365 files across the DDPs.
The two co-authors agreed on 336 files with a Fleiss' Kappa score of 0.85, indicating an almost perfect inter-annotator agreement between them~\cite{fleiss1971measuring}.
The third co-author then independently annotated the 29 files with disagreements to reach the final classification (based on majority agreement).
~\Cref{tab:ddp_categorization} presents an overview of our DDP categorization.

\subsection{Analyzing DDPs}

\begin{table}[t]
\caption{Number of \textit{Ad-only}, \textit{Non-ad}, and \textit{Mixed} files.}
\centering
\begin{tabular}{l *{3}{>{\centering\arraybackslash}p{1.7cm}} r}
\toprule
& \multicolumn{3}{c}{\textbf{\#Files across the three categories}} & \\
\cmidrule{2-4}
\textbf{Platform} & \textbf{Ad-only} & \textbf{Non-ad} & \textbf{Mixed} & \textbf{Total} \\
\midrule
Instagram & 7 & 21 & 51 & 79 \\
Facebook & 12 & 54 & 137 & 203 \\
X & 6 & 35 & 42 & 83 \\
\bottomrule
\end{tabular}
\label{tab:ddp_categorization}
\end{table}

Our analysis focuses on systematically comparing pairs of DDPs obtained from the same platform under different subscription models.
The comparison operates at multiple analytical levels.
At the file system level, we examine the overall structure of the DDPs, identifying differences in the presence, absence, or modification of files and directories.
At the data structure level, we traverse JSON files to detect changes in their schema and organization.
Together, these analyses allow us to capture both coarse and fine-grained differences in data collection practices across subscription models.

To support this process, we developed a browser-based analyzer tool that automated these comparisons and generated a structured JSON report summarizing the observed differences.
The tool is generalizable, as the implementation allows it to handle DDPs from multiple online platforms with almost no platform-specific customization.
We implemented the tool in JavaScript with client-side processing to ensure all data remains local to the user's machine.
To facilitate future research, we make our analyzer tool available.\footnote{\url{https://github.com/sepehrmousavi/subscription-data-practices}}

\subsection{Findings}
\label{subsec:rq1_findings}

\begin{table*}[t]
\caption{
Presence of Ad-only files of Instagram, Facebook, and X in both DDP-Ad-Based 1 and DDP-Ad-Free.
}
\centering
\begin{tabular}{>{\centering\arraybackslash}p{1.5cm}>{\raggedright\arraybackslash}p{10.0cm}>{\centering\arraybackslash}p{2.2cm}>{\centering\arraybackslash}p{2.2cm}}
\toprule
\textbf{Platform} & \multicolumn{1}{>{\centering\arraybackslash}p{10.8cm}}{\textbf{File}} & \textbf{Present in DDP-Ad-Based 1} & \textbf{Present in DDP-Ad-Free} \\
\midrule
\multirow{2}{*}[-0.5em]{Instagram} & \url{advertisers_using_your_activity_or_information.json}, \url{ads_about_meta.json}, \url{ads_clicked.json}, \url{ads_viewed.json} & \textbf{\fgcheck} & \textbf{\redtimes} \\

\cmidrule{2-4}
 
 & \url{other_categories_used_to_reach_you.json}, \url{subscription_for_no_ads.json}, \url{consents.json} & \textbf{\fgcheck} & \textbf{\fgcheck} \\

\midrule

\multirow{2}{*}[-3.5em]{Facebook} & \url{advertisers_using_your_activity_or_information.json}, \url{ads_about_meta.json}, \url{advertisers_you've_interacted_with.json}, \url{ads_interests.json}, \url{ads_personalization_consent.json} & \textbf{\fgcheck} & \textbf{\redtimes} \\ 

\cmidrule{2-4}

 & \url{ad_preferences.json}, \url{meta_ad_library_accounts.json}, \url{other_categories_used_to_reach_you.json}, \url{subscription_for_no_ads.json}, \url{your_consent_settings.json}, \url{your_meta_business_suite_guidance_interactions.json}, \url{consents.json} & \textbf{\fgcheck} & \textbf{\fgcheck} \\

\midrule

\multirow{1}{*}[-1em]{X} & \url{ad-engagements.js}, \url{ad-impressions.js}, \url{ad-mobile-conversions-attributed.js}, \url{ad-mobile-conversions-unattributed.js}, \url{ad-online-conversions-attributed.js}, \url{ad-online-conversions-unattributed.js} & \textbf{\fgcheck} & \textbf{\fgcheck} \\
\bottomrule
\end{tabular}
\label{tab:ddp_adonly}
\end{table*}

Here, we present findings from our analysis.
First, we begin with insights drawn from Ad-only files, then discuss Mixed files, and conclude with Non-ad files.

\smallskip
\noindent
\textit{\textbf{Ad-only files:}}
Ideally, once ads are no longer served, Ad-only files should demonstrate reduced data collection.
This may manifest either through their removal from the DDPs or by ceasing additional data collection during the ad-free subscription period.
To this end, we first examine whether Ad-only files are removed, and then assess whether the files that persist through both ad-based and ad-free periods exhibit reduced data collection.

~\Cref{tab:ddp_adonly} provides an overview of all Ad-only files from Instagram, Facebook, and X, and their presence in DDP-Ad-Based 1 and DDP-Ad-Free.
We observe that Instagram and Facebook removed 4 and 5 Ad-only files in their DDP-Ad-Free, corresponding to 57\% and 42\% of all Ad-only files in their DDPs, respectively, whereas X did not remove any.
This removal indicates that subscribing to the ad-free model stops Instagram and Facebook from processing the information included in these files.
To confirm, we compared DDP-Ad-Based 1 and DDP-Ad-Based 2 and found a lack of significant overlap between the two instances, suggesting that the contents of these files were indeed deleted after the ad-free subscription.

In addition, ideally, the remaining Ad-only files should cease data collection, particularly for user profiling.
To assess this, we examined whether such files continued collecting profiling-related data in DDP-Ad-Free.
Our analysis found that two Ad-only files on Meta platforms (one present in both Instagram and Facebook, and one unique to Facebook) and one Ad-only file on X continued to retain or collect profiling data.
These correspond to 14\%, 17\%, and 17\% of Ad-only files on Instagram, Facebook, and X, respectively.

In the case of Meta platforms, the file \textit{\url{other_categories_used_to_reach_you.json}}, present in both Instagram and Facebook, contains automatically inferred categories used by advertisers to target their audience~\cite{meta_categories_used_reach}.
We observed that this file retained its existing content in DDP-Ad-Free, but showed no signs of new data collection.
The retention of these categories allows Meta platforms to continue leveraging previous ad-related profiling data, even if they cease new data collection in the ad-free model.
Also, the file \textit{\url{ad_preferences.json}} in Facebook contains information related to advertising preferences.
This includes fields that have data on ads and advertisers hidden by the user, along with other user preferences.
Our analysis of DDP-Ad-Free revealed new entries in the hidden ads and hidden advertisers fields, indicating the propagation of profiling based on actions taken during the ad-based subscription period.
Conversely, we did not observe new data collection by the remaining Ad-only files in the Meta platforms.
Overall, these findings illustrate that, despite subscription to the ad-free model, Meta platforms continue to retain or collect some profiling data.

For X, the \textit{\url{ad-mobile-conversions-unattributed.js}} file stores mobile app events associated with an account that may become attributable to an advertisement in the future.
During the ad-free subscription period, this file recorded events (e.g., install and login) from two mobile applications, while the other Ad-only files showed no new data collection.
This indicates that X continues to collect exclusive ad-related profiling data even under the ad-free model.

\smallskip
\noindent
\textit{\textbf{Mixed files:}}
Mixed files contain data used for both advertising and non-advertising purposes (e.g., content personalization).
Hence, platforms are not necessarily required to alter their data collection practices for these files once users subscribe to the ad-free model.
Indeed, we did not find any such file being removed from DDP-Ad-Free across the three platforms.
However, Mixed files may include fields exclusively related to advertising, which ideally should no longer be processed once an account switches to the ad-free model.
To systematically assess this, we identified ad-related fields and examined whether they continued to be processed in DDP-Ad-Free.
To identify ad-related fields, we searched among Mixed files for keys containing the substring ``ad''.
This yielded 2 fields in Facebook and 5 in X, while none was found in Instagram.

In the case of Facebook, we identified \textit{advertiser\_id} and \textit{Google advertiser ID} fields inside \textit{\url{mobile_devices.json}} and \textit{\url{your_devices.json}} files, respectively.
Moreover, we observed that \textit{Google advertiser ID} was empty in all DDP instances.
However, \textit{advertiser\_id} retained its value in DDP-Ad-Free.
This identifier, which is associated with each mobile device, allows advertisers to target their audience and can be used to collect behavioral data of users~\cite{facebook_mobiler_adid}.
Although no ads are served under the ad-free model, the presence of this identifier still allows user tracking.

For X, we identified the \textit{audienceAndAdvertisers} field inside the \textit{\url{personalization.js}} file, which contains four nested ad-related fields: \textit{lookalikeAdvertisers}, \textit{advertisers}, \textit{doNotReachAdvertisers}, and \textit{catalogAudienceAdvertisers}.
These nested fields collectively contain information exclusively related to advertising purposes.
Moreover, they were all empty across DDP instances, except for \textit{lookalikeAdvertisers}.
In particular, \textit{lookalikeAdvertisers} in DDP-Ad-Free retained a non-empty list of advertisers that own look-alike audiences that include our X accounts.
Furthermore, these advertiser lists were identical between DDP-Ad-Based 1 and DDP-Ad-Free.
This raises the question of why X continued to include our accounts in look-alike audiences of advertisers despite the ad-free subscription.
It is essential to note that X provides advertisers with estimates of the ``reach'' (i.e., size) of their custom audiences.
If these estimates include a substantial number of ad-free users who cannot see ads, X risks misleading advertisers.

\smallskip
\noindent
\textit{\textbf{Non-ad files:}}
Naturally, subscribing to the ad-free model should not affect data collection practices of Non-ad files.
Indeed, our analysis confirms this expectation.
Specifically, we did not find any such file being removed from DDP-Ad-Free across the three platforms.
Moreover, we observed no meaningful or significant changes across the three platforms in how such files retain and collect data.
These observations suggest that the ad-free subscription model does not influence data collection practices of Non-ad files.

\smallskip
\noindent
\textbf{Main takeaways:} 

\smallskip
\noindent
\faHandPointRight
~Instagram and Facebook, upon subscription to the ad-free model, partially reduce their data collection practices by removing 57\% and 42\% of all Ad-only files in their DDPs, respectively, whereas X does not remove any such files (see~\Cref{tab:ddp_adonly}).

\smallskip
\noindent
\faHandPointRight
~However, we find evidence that all three platforms continue to retain or collect some personal data that are specifically categorized for advertising purposes under the ad-free model.

%% file: rq2.tex
\section{User Expectation on Data Gathering Practices Across Subscription Models}
\label{sec:rq2}

Having examined platform data collection practices, we next investigate user expectations by conducting a survey on perceptions of data collection under different subscription models.

\subsection{Survey Design}

Our survey comprised two parts.
The first examined users' normative and descriptive expectations regarding data collection practices across subscription models.
The second assessed whether differences or reductions in data collection affect participants' willingness to select an ad-free subscription.
The full questionnaire is provided in~\Cref{subsec:appendix_survey}, and we detail the survey design below.

\smallskip
\noindent
\textbf{Expectation of reduced data collection:}
First, we investigated participants' normative (what they believe platforms \textit{should} do) and descriptive (what they believe platforms \textit{would} do) expectations of data collection practices adopted by platforms based on the ad-based and ad-free subscription models.
To this extent, we asked our participants the following two questions:

\noindent
$\bullet$ \textit{Q\_Normative}:\textit{``If you subscribe to an ad-free plan of a platform, should the platform collect less data about you?''}.

\noindent
$\bullet$ \textit{Q\_Descriptive}:\textit{``If you subscribe to an ad-free plan of a platform, do you think that the platform would collect less data about you?''}.

We framed the above two questions as multiple-choice questions with the following three choices: (1) \textit{Yes}, (2) \textit{No}, and (3) \textit{I'm not sure}.
Furthermore, to account for potential order-effect bias caused by the ordering of these two questions, we 
randomly shuffled these two questions for each participant~\cite{perreault1975controlling}.
Also, to gain deeper insights into participants' reasoning, we placed a free-response question after each of these two questions.
Specifically, we asked our participants to justify their choice: 
\textit{``Please justify your above choice.''}

\smallskip
\noindent
\textbf{Choice of subscription model:}
The second part of our survey examined whether data collection practices of online platforms across subscription models might affect participants' willingness to choose an ad-free subscription.
We asked our participants:

\noindent
$\bullet$ \textit{Q\_Influence}:\textit{``Would it affect your decision to choose an ad-free plan if the platform continued collecting the same amount of data as it does in the ad-based plan?''}.

Similar to the first part of the survey, we also framed Q\_Influence as a multiple-choice question with the same three choices and accompanied it with a free-response question to understand the reasoning of our participants.

\subsection{Participant Recruitment}

We recruited three distinct user groups:
(1) legal experts,
(2) technology experts,
and (3) members of the general public.
Legal experts provide us with insights into normative and regulatory aspects of user privacy and data collection practices.
Technology experts offer informed views on what is technically feasible as well as what is being commonly implemented by online platforms.
Finally, members of the general public reflect end-user perceptions and concerns with respect to data collection practices.
By surveying these three groups, we aimed to understand how expectations vary across legal, technical, and end-user perspectives.

We used the online crowd-sourcing platform Prolific\footnote{\url{https://prolific.com}} and chose multiple screening criteria to recruit our participants.
To apply the above expertise-based distinction, we used Prolific’s ``Work Function'' screener, which identifies participants’ professional roles within their organizations.
This screener includes 21 work functions: we selected ``Legal'' for legal experts, ``Engineering (e.g., software)'' and ``IT / Information Networking / Information Security'' for technology experts, and the remaining 18 for members of the general public.
Since the legal basis of our work relies on the European digital regulation (e.g., GDPR), we targeted users who live in 
the European Union, where these regulations are applicable.
Moreover, as our questionnaire was in English, we recruited participants who were fluent in English.
We were interested in recruiting active and reliable participants.
Hence, we looked for users who had at least 20 Prolific submissions with an approval rate of at least 95\%.

We conducted our survey in May 2025.
The median survey completion time was $10$ minutes, and we reimbursed each participant with \textsterling 2, which is equivalent to \textsterling 12 per hour, more than the recommended remuneration rate by Prolific.\footnote{\url{https://researcher-help.prolific.com/en/article/9cd998}}
We aimed to recruit 100 participants for each of the three user groups.
While we successfully reached this target for both tech experts and members of the general public, we were only able to recruit 85 legal experts, resulting in a total of 285 participants.
This shortcoming was due to the limited number of Prolific users who satisfied our screening requirements.
In particular, only 190 active Prolific users met our screening criteria for the legal experts group.
Therefore, our recruitment capacity was constrained for this specific user group.
We provide demographic characteristics of our participants in~\Cref{subsec:appendix_demo}.

\smallskip
\noindent
\textbf{Ethical Considerations:}
The Ethical Review Board (ERB) of Saarland university approved our study.
Participants took part in our survey voluntarily and with informed consent.
Furthermore, pursuant to the ERB guidelines, we did not collect any personally identifiable information (PII) from our participants.

\subsection{Survey Validation and Qualitative Coding of Free-response Questions}

We improved our survey after performing pilots with two of our colleagues and obtaining their feedback.
Then, we conducted three pilot studies on Prolific, each with 40 participants.
This is a much higher sample size than the recommended baseline for survey validation~\cite{perneger2015sample}.
We ensured that our pool of participants in these pilot studies differed from that in our final study.

We used qualitative open coding methodology to analyze the free-response questions and identify recurring themes in participants' explanations.
We grouped responses by their corresponding multiple-choice answer (Yes, No, or I'm not sure), resulting in nine sets.
Then, the first author of this work reviewed and coded these sets.
We provide further details and examples in~\Cref{subsec:appendix_coding}.

Furthermore, in line with prior work, we relied on answers to the three free-response questions in our survey to detect participants who were inattentive~\cite{farke2023does, dash2024investigating}.
We compared each participant’s multiple-choice and free-response answers, and flagged a participant as inattentive if the free-response answer (1) did not match the multiple-choice answer, (2) was missing, or (3) contained only a couple of words.
In the end, we found 30 inattentive participants and discarded their responses.
Hence, we analyzed the responses of the remaining 255 participants, consisting of 78 legal experts, 90 technology experts, and 87 members of the general public.

\subsection{Survey Results}

This subsection reports the results of our descriptive survey which aims to quantify the gap between users’ normative and descriptive expectations.
Hence, we focus on characterizing these distributions at the population level and do not conduct inferential statistical analyses to support our main claims.

\begin{figure*}[t]
\centering
    \begin{subfigure}[t]{0.33\textwidth}
        \centering
        \includegraphics[width=\linewidth]{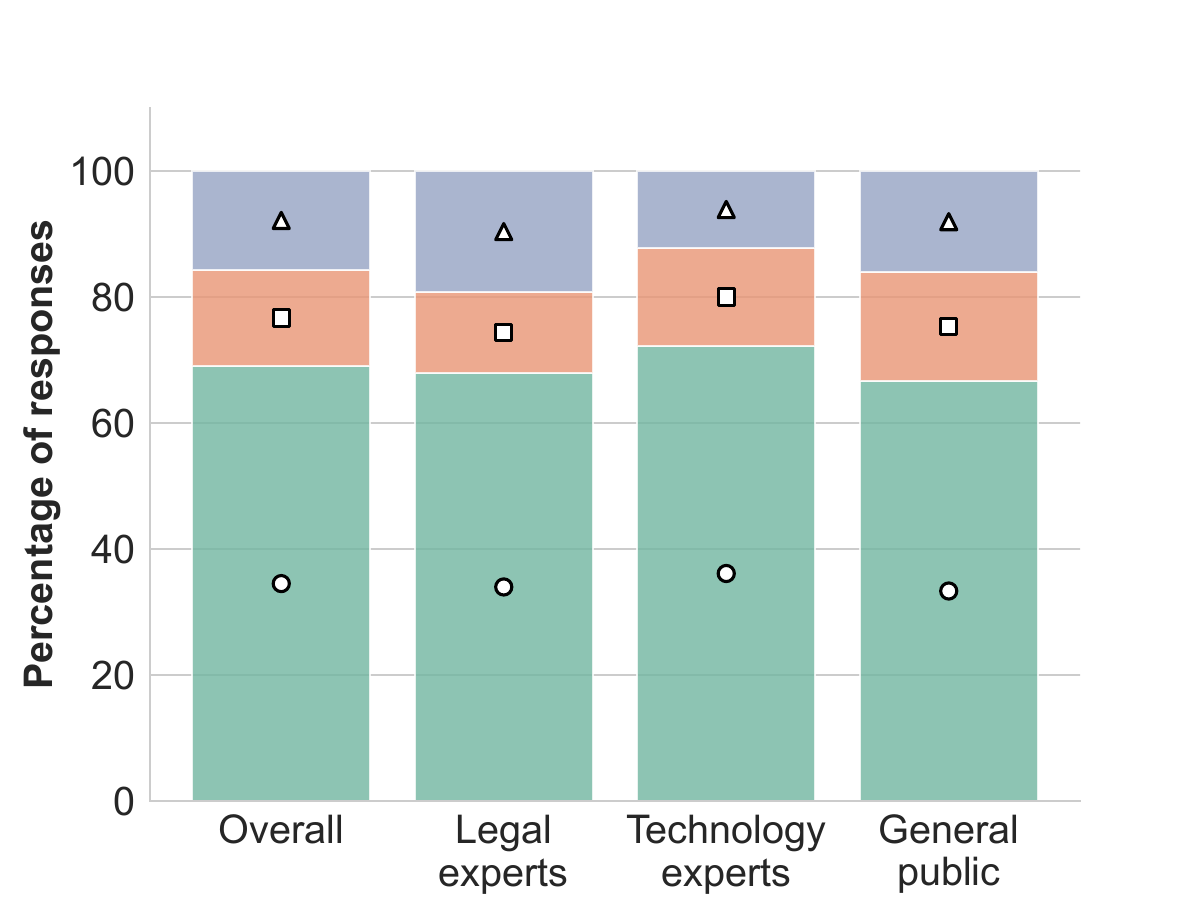}
        \caption{Q\_Normative}
        \label{fig:normative}
    \end{subfigure}
    \hfill
    \begin{subfigure}[t]{0.33\textwidth}
        \centering
        \includegraphics[width=\linewidth]{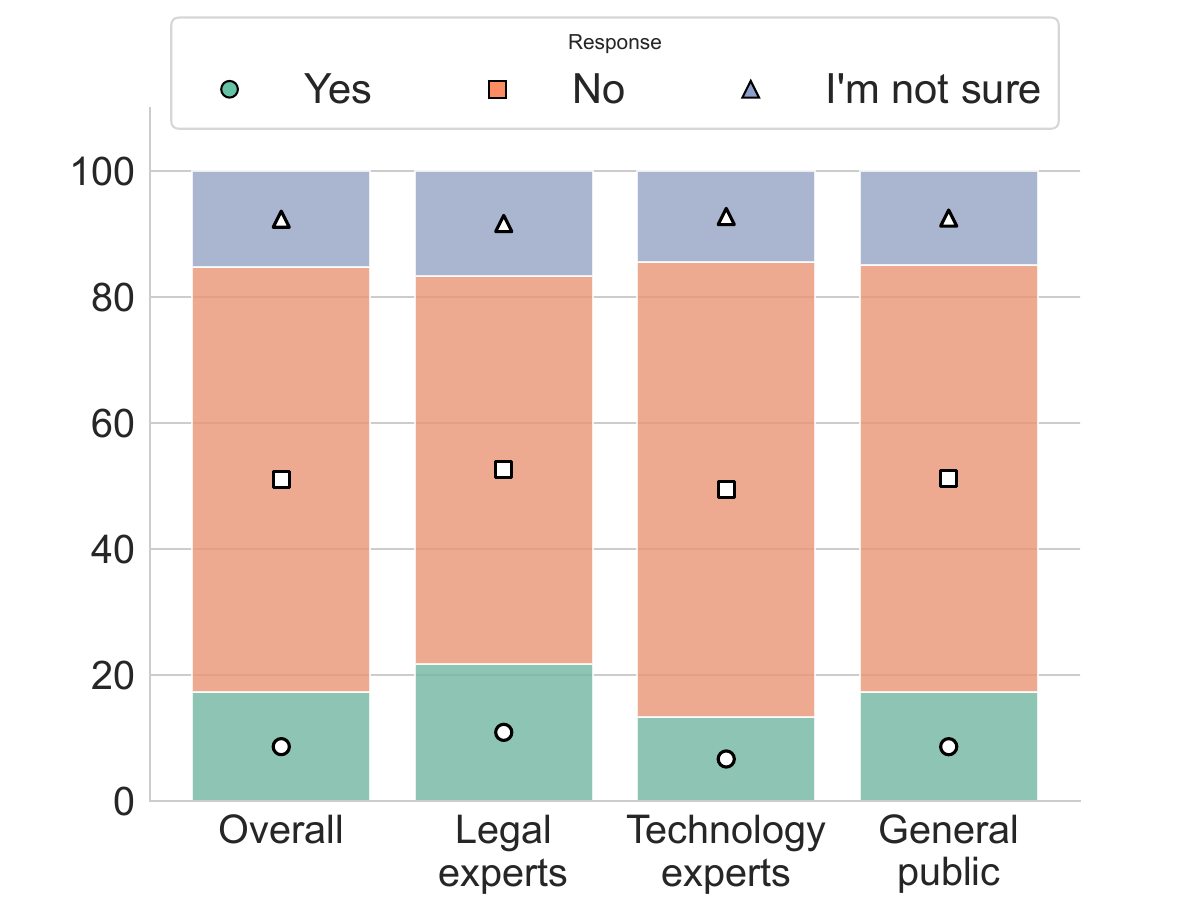}
        \caption{Q\_Descriptive}
        \label{fig:descriptive}
    \end{subfigure}
    \hfill
    \begin{subfigure}[t]{0.33\textwidth}
        \centering
        \includegraphics[width=1\columnwidth]{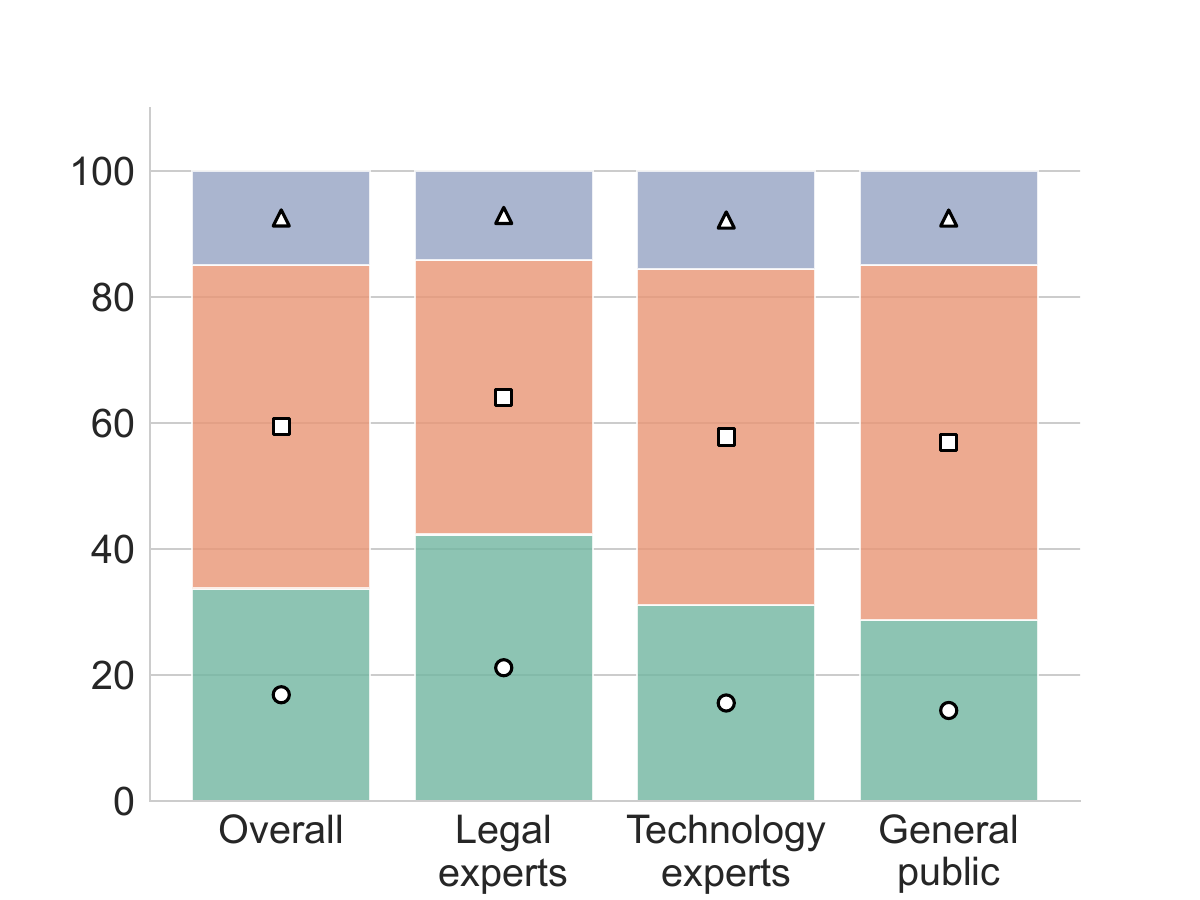}
        \caption{Q\_Influence}
        \label{fig:influence}
    \end{subfigure}%
    \caption{
    Distribution of participants' responses to Q\_Normative, Q\_Descriptive, and Q\_Influence.
    We observe that, overall, 69\% of participants believed that online platforms should collect less user data in the ad-free model compared to the ad-based model. However, only 17\% of participants believed that online platforms would actually reduce their data collection practices in reality.
    Furthermore, 34\% of participants stated that the continued collection of the same amount of data in the ad-free model as in the ad-based model would deter their decision to subscribe to the ad-free model.
    }
    \label{fig:norm_descript_influence}
\end{figure*}

\smallskip
\noindent
\textbf{Normative expectation of reduced data collection:}
~\Cref{fig:normative} shows the distribution of responses to Q\_Normative.
We observe that, overall, 69\% of participants believed that online platforms should collect less data from users subscribed to the ad-free model compared to the ad-based model.
Moreover, we see that the normative expectations of participants are closely aligned across the three user groups.
Specifically, 68\% of legal experts, 72\% of technology experts, and 67\% of the general public responded positively to Q\_Normative.
The substantial majority of respondents who normatively expect less data collection also signal a desire for stronger digital privacy beyond just ad removal.
This highlights a clear gap between users’ normative expectations and the actual platform practices observed in~\Cref{sec:rq1}.

To further understand the reasoning of participants, we analyzed their answers to the free-response question that followed Q\_Normative.
Since the normative expectations of participants are closely aligned across the three user groups, we combined the answers to the free-response question across the three user groups.
Among the participants who responded positively to Q\_Normative, 37\% ($n=65$) mentioned that \textit{data is no longer needed for targeted advertising}, 28\% ($n=49$) stated that \textit{ad-free subscription payment should stop monetization of their data}, and 13\% ($n=23$) believed that \textit{privacy should prevail as a benefit of paying the subscription fee}.

Among the participants who responded negatively (No) to this question, 28\% ($n=11$) thought that \textit{data collection is needed for other platform functionalities beyond advertising}, 23\% ($n=9$) mentioned their \textit{indifference to data collection}, and 21\% ($n=8$) stated that \textit{the scope of the ad-free subscription only covers ad removal}.

\smallskip
\noindent
\textbf{Descriptive expectation of reduced data collection:}
Interestingly, although the majority of participants normatively expect a reduction in data collection, only 17\% of our participants believed that online platforms \textit{would} actually collect less data from users subscribed to the ad-free model compared to the ad-based model (see~\Cref{fig:descriptive}).
Furthermore, we see that this expectation of participants is comparable across the three user groups.
Particularly, 22\% of legal experts, 13\% of technology experts, and 17\% of general public responded positively to Q\_Descriptive.

Importantly, among the participants who normatively expect platforms to reduce data collection, 63\% expressed doubt that platforms would actually reduce data gathering.
This skepticism aligns with the observed data collection practices of platforms (see~\Cref{sec:rq1}) and highlights users’ limited trust in online services.

Next, we analyzed participants’ answers to the free-response question following Q\_Descriptive by aggregating responses across the three user groups.
Among the participants who responded positively (Yes) to Q\_Descriptive, 55\% ($n=24$) mentioned that \textit{data is no longer needed for serving ads} and 20\% ($n=9$) stated that \textit{ad-free subscription payment should stop monetization of their data}.

Moreover, among the participants who responded negatively (No) to Q\_Descriptive, 38\% ($n=65$) said that \textit{data collection is profitable for platforms}, 21\% ($n=36$) believed that \textit{data collection is a standard practice of platforms}, and 13\% ($n=23$) thought that \textit{the scope of the ad-free subscription only covers ad removal}.

\smallskip
\noindent
\textbf{Choice of subscription model: }
~\Cref{fig:influence} shows the distribution of participants' responses to Q\_Influence. 
Overall, 34\% of participants mentioned that continued collection of the same amount of data would deter their decision to choose an ad-free plan.
Moreover, we see that the prevalence of this belief is quite comparable across the three user groups.
Specifically, 42\% of legal experts, 31\% of technology experts, and 29\% of the general public responded positively
to Q\_Influence.

To understand the reasoning of participants, we analyzed their answers to the free-response question following Q\_Influence by aggregating responses across the three user groups.
Among the participants who responded positively (Yes) to Q\_Influence, 43\% ($n=37$) mentioned that \textit{paying for an ad-free subscription is not financially worth it because platforms still collect the same amount of data} and 30\% ($n=26$) stated \textit{privacy and data protection concerns}.
Furthermore, among the participants who responded negatively (No) to Q\_Influence, 30\% ($n=39$) stated that \textit{data collection is an inevitable practice of online platforms}, 28\% ($n=37$) primarily wanted to \textit{only avoid viewing ads}, and 27\% ($n=35$) raised \textit{their indifference and lack of concern}.

\smallskip
\noindent
\textbf{Main takeaways:}

\smallskip
\noindent
\faHandPointRight~The majority of surveyed participants (69\%) normatively expect platforms to reduce data collection in ad-free model.

\smallskip
\noindent
\faHandPointRight~However, only 17\% of our participants descriptively expect platforms to reduce data collection in ad-free model.

\smallskip
\noindent
\faHandPointRight~Furthermore, 34\% of participants mentioned that the continued collection of the same amount of data in the ad-free model as in the ad-based model would deter their decision to subscribe to the ad-free plan.

%% file: discussion.tex
\section{Concluding Discussion}

\subsection{Summary}

We studied how platforms differ in their data collection practices across different subscription models and how users perceive the data collection practices associated with these subscription models.
Through exercising GDPR \textit{Right of access} (Article 15(3)), we empirically studied the data collection practices of Instagram, Facebook, and X across their ad-based and ad-free subscription models.
Our analysis revealed some evidence that Instagram and Facebook reduced their data collection by removing certain ad-related data under the ad-free model.
However, we observed that all three platforms continue to retain or collect some personal data that we classify as advertising-related, even under the ad-free model.
Next, by surveying 255 participants, we investigated user perception of data collection practices.
We found that 69\% of our participants normatively believe that platforms should reduce data collection in ad-free subscription model.
Whereas, only 17\% of them expect platforms to reduce their data collection in practice.

\subsection{Limitations}

Despite its contributions, our study has some limitations.
First, our study is limited to two author-owned accounts on three platforms, which cannot capture the full range of user behaviors (e.g., device diversity or feature usage) and constrains the robustness of our findings across platforms.
However, this restriction is deliberate, as analyzing third-party users' DDPs would require processing sensitive PII and creating multiple fake accounts would violate platforms' terms of service~\cite{Facebook_multiaccount}.
Hence, our findings establish a lower bound on what data continues to be collected or retrained in ad-free models.
Also, to evaluate robustness across platforms, we already analyzed YouTube.
However, we excluded it as its DDP doesn’t include any ad-related information other than viewed ads, which prevents us from drawing comparable conclusions.
Second, although GDPR Article 15(3) mandates comprehensive data disclosure, the completeness of DDPs cannot be independently verified due to the proprietary nature of such practices. 
Hence, our analysis is limited to the data available within these DDPs and our findings should be interpreted within the visibility scope of DDPs.

\subsection{Discussion}

Despite the above limitations, our insights reveal a clear gap between user demands and platform behavior.
While 69\% of surveyed participants believe platforms should collect less data under the ad-free model, our empirical analysis shows that platforms continue to retain or collect certain data that are processed exclusively for advertising purposes.
This mismatch means that users are not receiving what they believe they are paying for, as subscribing to an ad-free model does not translate into reduced data collection.
Notably, 34\% of participants reported that the continued collection of the same amount of data in the ad-free model as in the ad-based model would deter their decision to pay for the ad-free subscription.
This observation highlights how this gap undermines user trust and risks misinterpretation of subscription offerings.

Furthermore, our findings point to two broader implications.
First, data collection practices under the ad-free model raises concerns about compliance with core GDPR principles, such as purpose limitation and data minimization (Article 5(1)).
This calls for closer regulatory scrutiny of whether platforms meaningfully adjust their data practices across subscription models, alongside enforcement efforts that help ensure that ``ad-free'' reflects a substantive shift rather than a rhetorical label.
Second, our results indicate at potential internal misalignment within platforms themselves.
Even when companies aim to provide more transparency or comply with regulation, technical and organizational silos may prevent the implementation of fully-compliant solutions.
Moving forward, we believe that there is a need for more coordination across legal, engineering, and product teams to ensure that platform-wide data practices align with both user demands and regulatory requirements.

%% file: appendix.tex
\appendix
\label{sec:appendix}

\section{Demographic statistics of survey participants}
\label{subsec:appendix_demo}

\begin{table}[h]
\caption{Demographic distribution of survey participants.}
\centering
\begin{tabular}{@{}lcrr@{}}
\toprule
\textbf{Characteristic}                & \textbf{Category}     & \multicolumn{1}{c}{\textbf{Count}} & \multicolumn{1}{c}{\textbf{Percentage}} \\ \midrule
\multirow{2}{*}{\textbf{Gender}}  & Male              & 167                                & 65.5                                     \\
                                  & Female            & 88                                & 34.5                                   \\ \midrule
\multirow{5}{*}{\textbf{Age}}     & 18-24             & 41                                 & 16.1                                    \\
                                  & 25-34             & 128                                & 50.2                                      \\
                                  & 34-44             & 47                                & 18.4                                   \\
                                  & 45-64             & 38                                & 14.9                                    \\ 
                                  & 65+               & 1                                 & 0.4                                    \\ \bottomrule
\end{tabular}
\label{tab:demographics}
\end{table}

~\Cref{tab:demographics}
summarizes two main demographic characteristics of our participants, namely gender and age.
We observe that the gender of our recruited participants is skewed towards Male users.
Furthermore, we see that the age distribution of our participants is skewed towards users who are young and at most 34 years old.


\section{Information contained in the DDPs of Instagram, Facebook, and X}
\label{subsec:appendix_ddp}

~\Cref{ddp_table} contains detailed descriptions of the fields of information made available in the DDPs of Instagram, Facebook, and X.

\begin{table*}[t]
\centering
\caption{Fields of information available in the data download packages of Instagram, Facebook, and X, along with their descriptions.}
\begin{tabular}{p{0.20\linewidth} p{0.40\linewidth} p{0.35\linewidth}}
\toprule
\textbf{Online Platform} & \textbf{Field of Information} & \textbf{Description} \\
\midrule
Instagram/Facebook 
& \textbf{Your Instagram/Facebook activity} 
& Information about user activity, such as content created, messages and more \\
& \textbf{Personal information} 
& Information about user profile \\
& \textbf{Connections} 
& Who and how user has connected with people \\
& \textbf{Logged information} 
& Information that Meta logs about user activity, including things like user location and search history \\
& \textbf{Security and login information} 
& Technical information and login activity \\
& \textbf{Apps and websites off of Instagram/Facebook} 
& Apps user owns and activity Instagram/Facebook receives from apps and websites off of Instagram/Facebook \\
& \textbf{Preferences} 
& Actions user has taken to customize their experience \\
& \textbf{Ads information} 
& User interactions with ads and advertisers \\
\midrule
X (formerly Twitter)
& \textbf{Sensory information} 
& Audio, electronic, visual, and similar information \\
& \textbf{Identifiers} 
& Real name, alias, postal address, telephone number, unique identifiers (such as a device identifier, cookies, mobile ad identifiers), customer number, Internet Protocol address, email address, account name, and other similar identifiers \\
& \textbf{Online activity} 
& Internet and other electronic network activity information, including, but not limited to, information regarding interactions with websites, applications, or advertisements \\
& \textbf{Inferences} 
& Inferences drawn to create a profile about the user reflecting their preferences, characteristics, predispositions, behavior, and attitudes \\
& \textbf{Protected classifications} 
& Characteristics of certain legally protected classifications \\
\bottomrule
\end{tabular}
\label{ddp_table}
\end{table*}


\section{Survey questions}
\label{subsec:appendix_survey}

In this section, we provide our questionnaire.

\begin{quote}
\itshape
This section focuses on your views regarding how data collection practices \textbf{should} differ between a platform’s ad-based and ad-free subscription plans.
\end{quote}

\begin{enumerate}
    \item If you subscribe to an ad-free plan of a platform, \textbf{should} the platform collect less data about you?
    \begin{itemize}
        \item Yes
        \item No
        \item I'm not sure
    \end{itemize}

    \item Please justify your above choice.

\end{enumerate}

\begin{quote}
\itshape
This section asks \textbf{what you think the potential differences will be} in user data collection between ad-based and ad-free subscription plans on a platform.
\end{quote}

\begin{enumerate}
    \item If you subscribe to an ad-free plan of a platform, \textbf{do you think} that the platform \textbf{would} collect less data about you?
    \begin{itemize}
        \item Yes
        \item No
        \item I'm not sure
    \end{itemize}

    \item Please justify your above choice.

\end{enumerate}

\begin{quote}
\itshape
This section focuses on whether continued collection of the same amount of data between ad-based and ad-free subscription plans affects your willingness to choose an ad-free subscription plan.
\end{quote}

\begin{enumerate}
    \item Would it affect your decision to choose an ad-free plan if the platform continued collecting the same amount of data as it does in the ad-based plan?
    \begin{itemize}
        \item Yes
        \item No
        \item I'm not sure
    \end{itemize}

    \item Please justify your above choice.

\end{enumerate}


\section{Qualitative coding of free-response questions}
\label{subsec:appendix_coding}

For each free-response question, we grouped responses according to the corresponding multiple-choice answer (Yes, No, or I’m not sure), resulting in nine sets in total.
Then, one co-author independently reviewed each set and identified all the broad themes that captured ideas expressed in the responses.
We present the three most common themes for each set, along with a random participant response for each theme in~\Cref{tab:qualcoding}.

\begin{table*}[t]
\centering
\caption{The three most common themes observed for each set after conducting qualitative coding of free-response answers.}
\begin{tabular}{p{2.0cm} p{3.5cm} p{4.0cm} p{7.5cm}}
\toprule
\textbf{Question} & \textbf{Multiple-choice answer} & \textbf{Coded free response} & \textbf{Example free-response answer} \\

\midrule

\multirow{9}{*}{Q\_Normative} & \multirow{3}{*}{Yes} & Reduced need (ad-free) & \textit{They don't need data for personal advertising, so they could stop collecting them.} \\

\cmidrule{3-4}

 &  & Payment principle & \textit{If something is free, you are the product.} \\

\cmidrule{3-4}

 &  & Privacy expectation & \textit{If I am paying for a service it should come with benefits in terms of privacy as well} \\

\cmidrule{2-4}

 & \multirow{3}{*}{No} & Data needed (other functions) & \textit{They need data to provide other features} \\

\cmidrule{3-4}
 
 &  & User indifference & \textit{Dont care} \\

\cmidrule{3-4}
 
 &  & Scope of ad-free plan & \textit{I think you're just paying to not have ads.} \\
 
\cmidrule{2-4}

 & \multirow{3}{*}{I'm not sure} & Lack of knowledge & \textit{i don't understand these data things} \\

\cmidrule{3-4}

 &  & Depends on specifics & \textit{Maybe it just depends on the agreement.} \\

\cmidrule{3-4}

 &  & Skepticism & \textit{I cant control the data collected by the platform, but probably they collect every single click on ad-free or ad-based} \\
 
\cmidrule{1-4}

\multirow{9}{*}{Q\_Descriptive} & \multirow{3}{*}{Yes} & Reduced need (ad-free) & \textit{Because without ads, less data is needed for targeting and personalization.} \\

\cmidrule{3-4}

 &  & Payment changes incentive & \textit{Since I’m paying and the platform is making a profit, why does it still collect so much data?} \\

\cmidrule{3-4}
 
 &  & Expected benefit of plan & \textit{a premium service has certain benefits and I think this was the case} \\

\cmidrule{2-4}

 & \multirow{3}{*}{No} & Data value & \textit{The payment for the service are data} \\

\cmidrule{3-4}
 
 &  & Standard practice & \textit{I think they still collect the same data as before.} \\

\cmidrule{3-4}

 &  & Scope of plan & \textit{I don't think it matters to them, ad free means just that, not "more privacy".} \\
 
\cmidrule{2-4}

 & \multirow{3}{*}{I'm not sure} & Lack of knowledge & \textit{I honestly am not very informed about the data collection on ad-free platforms} \\

\cmidrule{3-4}

 &  & Skepticism & \textit{I'm not sure as I don't really trust how the platforms collect our data} \\

\cmidrule{3-4}

 &  & Conflicting beliefs & \textit{I am thinking it twice maybe they collect my personal data but I don't see relevant ads} \\
 
\cmidrule{1-4}

\multirow{9}{*}{Q\_Influence} & \multirow{3}{*}{Yes} & Value proposition & \textit{If I know it will collect more or the same amount, I just might refuse the service as it would not be worth it} \\

\cmidrule{3-4}

 &  & Privacy expectation & \textit{Because one of the main reasons to pay for ad-free is to reduce data collection and protect privacy.} \\

\cmidrule{3-4}
 
 &  & Reduced need (ad-free) & \textit{if there's no ad, they dont need my data} \\

\cmidrule{2-4}

 & \multirow{3}{*}{No} & Inevitability & \textit{It is so universal that it is impossible to fight anymore} \\

\cmidrule{3-4}
 
 &  & Focus on ad removal & \textit{Because when buying such a package I am primarily interested in the lack of advertisements} \\

\cmidrule{3-4}

 &  & User indifference & \textit{I just wouldnt care about it, as long as I get no ads} \\
 
\cmidrule{2-4}

 & \multirow{3}{*}{I'm not sure} & Depends on specifics & \textit{Depends. I'm using adblocker everywhere, so ad-free plans don't have a lot of value for me} \\

\cmidrule{3-4}

 &  & Lack of knowledge & \textit{not sure have no knowledge in this really} \\

\cmidrule{3-4}

 &  & Focus on ad removal & \textit{I would probably use the plain mainly to not see ads, the data collection is secondary} \\

\bottomrule
\end{tabular}
\label{tab:qualcoding}
\end{table*}